# Effect of SW defect on structural and transport properties of silicene nanoribbons

Dace Zha[a], Changpeng Chen[a,*], Manman Wang[a]

School of science, Wuhan University of Technology, Wuhan 430070 PR China

**Abstract:** Using density functional theory and non-equilibrium Green's function technique, we performed theoretical investigations on the structural and transport properties of zigzag silicene nanoribbons (SiNRs) with Stone-Wales defect. The calculated formation energy is significantly lower than that of graphene and silicene, which implies the high stability of such defect in SiNRs. Negative differential resistance (NDR) can be observed within certain bias voltage range in both perfect and SW defected SiNRs. In order to elucidate the mechanism the NDR behavior, the transmission spectra and molecular projected self-consistent Hamiltonian (MPSH) states are discussed in details.

## 1. Introduction

Silicon is an element next to carbon among the group-IV elements which also has stable mono-layer honeycomb structure. Silicene, a graphene analogue for silicon, which is predicted to be a zero band gap semiconductor with a Dirac cone as observed in graphene, is rapidly attracting significant interest in the basic and engineering sciences [1-5]. Different from the famous graphene, both theoretical and experimental studies show that silicene is not planar but has a low-buckled honeycomb structure with buckled height of about 0.44Å for stability [6-9]. The underlying reason is that silicene is energetically more favorable as a buckled structure since its sp3 hybridization is more stable than its sp2 form [10,11].

Recently, many progresses have been made in the field of silicene based materials, for example, silicene nanoribbons (SiNRs) were synthesized through epitaxial growth of silicon on the substrate of Ag(110)[12,13]. However, most of the experimentally synthesized sheets comprise different Stone–Wales (SW) defects. These defects have been known to result in dramatic changes of their electronic and mechanical properties both in graphene [14-17] and silicene [18-19]. X.F.Yang et al [20] investigated the spin-polarized transport properties of the bare and hydrogenated zigzag silicene nanoribbons (ZSiNRs) and find that the negative differential resistances exists in ZSiNRs. Jun Kang et al [21] studied the transmission spectra for 6-ZSiNR and 7-ZSiNR under a small bias voltage and find Even-N and odd-N ZSiNRs have different current-voltage relationships. While, the effects of Stone–Wales defects on the transport properties of silicene nanoribbons (SiNRs) have remained unexplored. In this article, we focus on the structural and transport properties of perfect and SW defected zigzag SiNRs with first principle calculation. Our purpose is to investigate the effect of the defects on the structural and transport properties of SiNRs theoretically. The paper is organized as follows: Section 2 describes the model and computational methodology. Section 3 discusses the structural change and stability of SW defects in SiNRs. Section 4 presents the transport properties of defected and defect-free SiNRs.

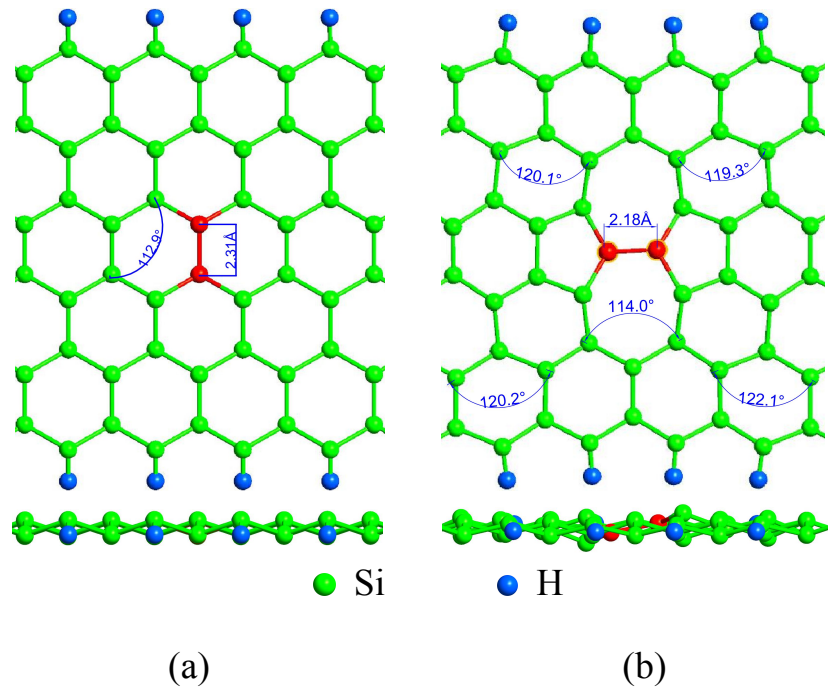

Fig. 1. The optimized structure of (a) perfect and (b) SW defect cases.

## 2. Model and method

The optimized geometric structures of the SiNRs with and without SW-defect from the top view (a) and from the side view (b) are illustrated in Fig. 1. A SW defect is created by the rotation of a silicon dimer by 90º around the center of the Si-Si bond. The structures were optimized and the quantum transport calculations were carried out by the Atomistix ToolKit (ATK) package [22-23], which bases on the fully self-consistent non-equilibrium Green's functions and the density functional theory. Under external bias, the current through a molecular junction is calculated from the Landauer–Büttiker formula:

$$I(V) = 2e/h \int [f(E-\mu_L) - f(E-\mu_E)] T(E,V) dE$$

Where f is the Fermi–Dirac distribution for the left (L) and right (R) electrodes, $\mu_{L/R}$ the electrochemical potential of the left or right electrode and the difference in the electrochemical potential is given by eV with the applied bias voltage V. $\mu_L=\mu(0)-eV/2$ and $\mu_R=\mu(0)+eV/2$. Furthermore, $\mu_{L/R}(0)=E_F$ is the FL. The $T(E,V)=T_R[\Gamma_L(E,V)G^R(E,V)\Gamma_R(E,V)G^A(E,V)]$ is the transmission spectrum, where $G^{R/A}$ are the retarded and advanced Green's functions, and coupling functions $\Gamma_{L/R}$ are the imaginary parts of the left and right self-energies, respectively. Self energy depends on the surface Green's functions of the electrode regions and comes from the nearest-neighbor interaction between the extended molecule region and the electrodes. For the system at equilibrium, the conductance G is evaluated by the transmission function T(E) at the Fermi level (FL) $E_F$ of the system: $G=G_0T(E_F)$, where $G_0=2e^2/h$ is the quantum unit of conductance, and h is Planck's constant and e is the electron charge.

The geometric optimization and transport properties are calculated by ATK within local density approximation for exchange-correlation potential. The exchange correlation functional is set as local-density approximation (LDA) in the PerdewZunger (PZ) form. The mesh cutoff is set to be 150 Rydberg. The structure of the two-probe system of perfect and SW defected SiNRs are illustrated in the Fig. 2. We calculate the electronic structure of the semi-infinite electrodes with a Monkhorst-Pack grid of 1×1×100 k-point. The basis set is adopted for elements of systems and the convergence criterion for the total energy is $10^{-4}$ Ry.

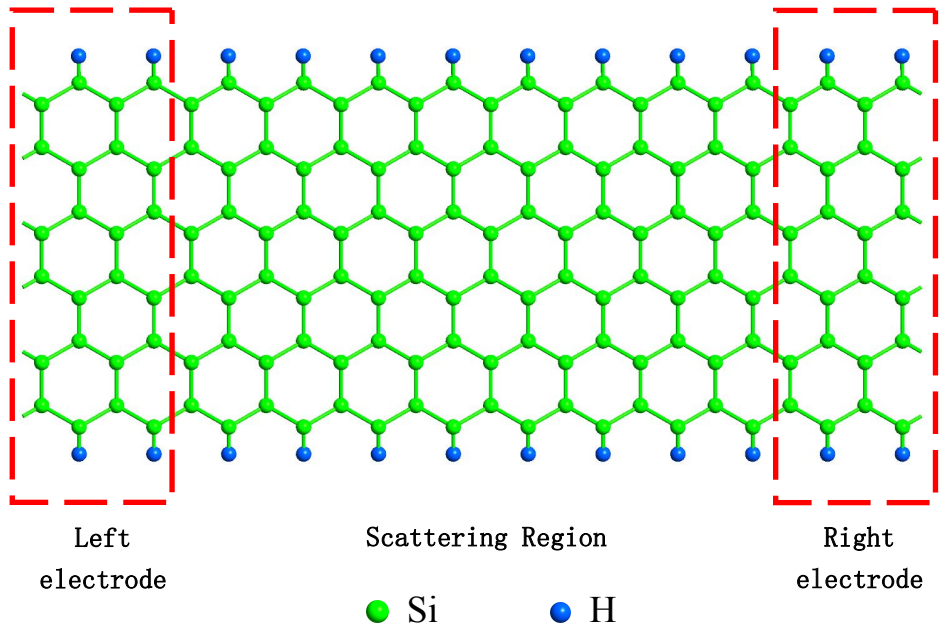

(a)

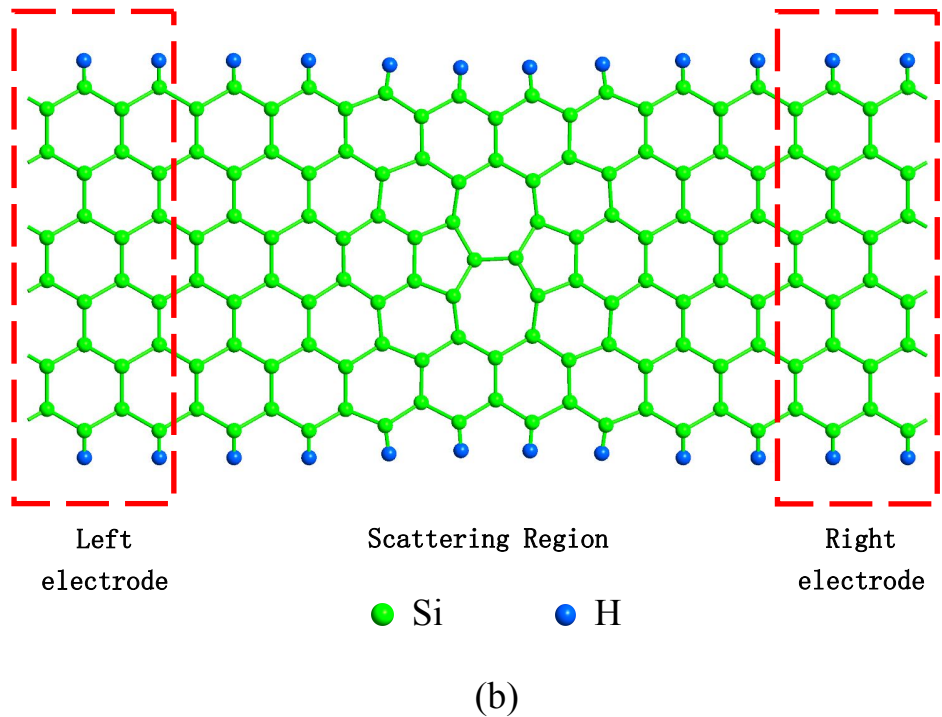

Fig.2 The structure of the two-probe system of (a) perfect SiNRs (b) SW defected SiNRs

## 3.1 The stability of SW defect in SiNRs

After the formation of the SW defect, four neighboring hexagons of perfect SiNRs are transformed into a pentagon and a heptagon pair. As illustrated in Fig. 1, the length of Si—Si bond which links two heptagons is 2.18A, shorter than its counter part in defect-free SiNRs. The average length of the two pentagonal bonds and the two heptagon bonds are 2.257Å, 2.267Å respectively. Corresponding to the length change of bonds, the bond angles of the adjacent hexagon change too. The biggest bond angles adjacent to the pentagons and heptagons are120.1º and 119.3º，respectively, differ with perfect silicon material 7.2º and 6.4º, respectively. The large variation of the bond lengths and bond angles destroys the symmetry of the whole system. It can also be seen that the SW defected SiNRs still has a low-buckled honeycomb structure, while the two is atoms, which marked in red in Fig.1, their extent of convex and concave decreases. The causes of such changes are partly due to the introduction of the SW defect, others owing to the thermal fluctuation of the system.

To investigate the stability of SW defect in SiNRs, we perform the calculation of the formation energy. The formation energy $E_f$ of SW-defect SiNRs is given by $E_f = E_{sw} - E_{perfect}$, where $E_{sw}$ and $E_{perfect}$ are respectively the total energies of the system with and without SW defect. The energies of Perfect SiNRs and SW-defected SiNRs are calculated to be -17326.9 eV and -17326.4 eV, respectively. $E_f$ for the SiNRs was found to be 0.5eV, much lower than that of graphene [24] (4.66-5.63eV) and the infinite silicene sheet [25] (1.64-1.82eV), which implies the higher stability of such defects in SiNRs than in graphene and silicene.

**4. The transport properties of perfect and SW defected SiNRs**

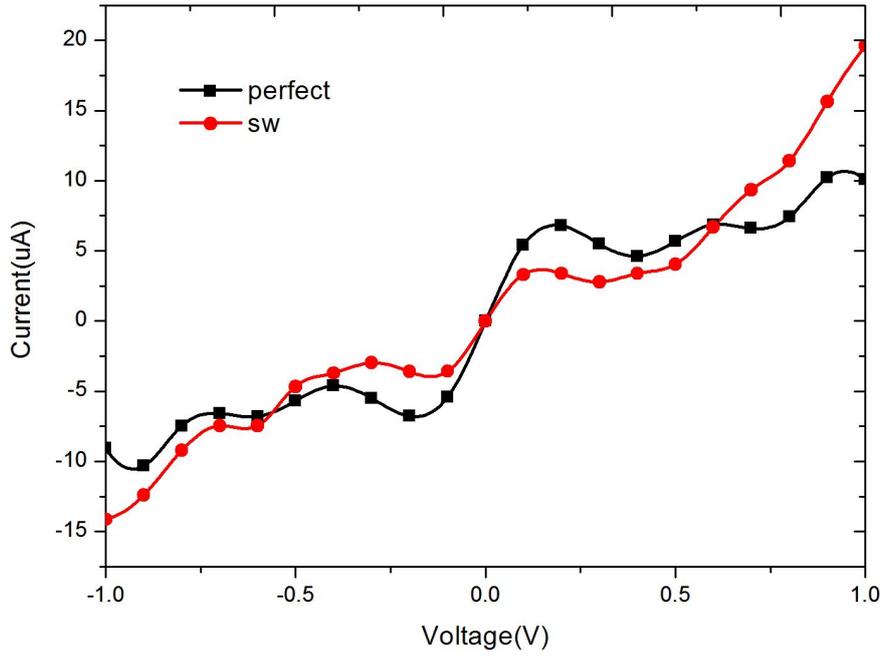

Fig. 3 The currents as functions of the applied bias for the perfect SiNRs and the SW-defected SiNRs

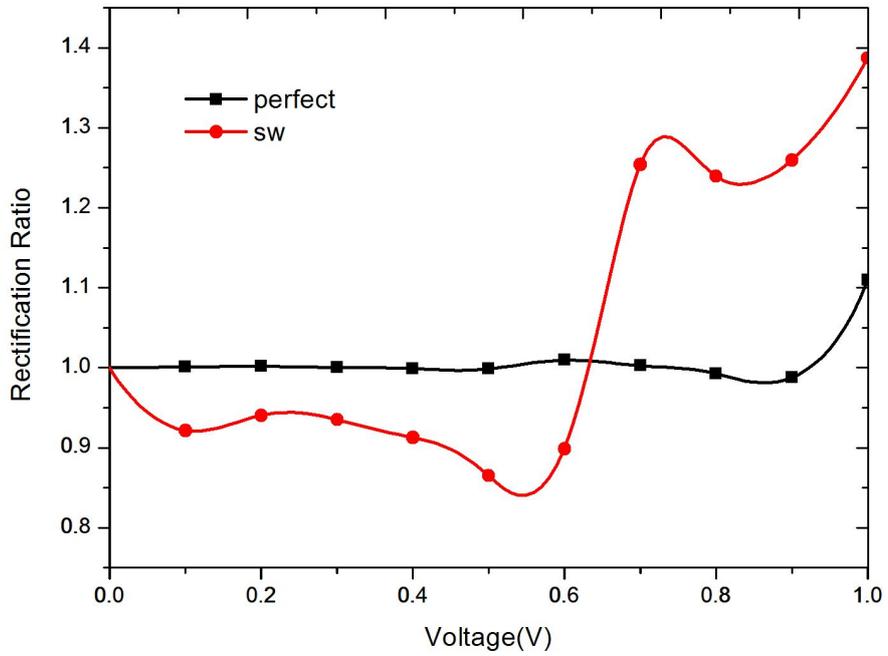

Fig. 4, The rectification ratio as a function of applied bias voltage for SW-defected SiNRs and perfect SiNRs.

In Fig. 3, we describe the I-V characteristic for the perfect and SW-defected SiNRs from -1 to 1 V in steps of 0.1V, respectively. It can be seen that the I-V curve of perfect SiNRs show obviously asymmetric behavior. When |V| varies from 0 V to 0.3V, the current increases quickly at positive voltage, attains to the maximum at the positive voltage of 0.2 V for the perfect SiNRs structure. One can clearly see that the current of perfect SiNRs decreases rapidly with increasing voltage, showing a negative differential resistance (NDR) phenomenon. Interestingly, the NDR behaviors can be observed in the I-V curves for SW-defected SiNRs at lower bias ranges too, and the characteristics of NDR phenomenon of perfect and defected SiNRs are similar from 0V to 0.3V. The difference appears when V>0.3V, where two peaks appear in the curve of perfect SiNRs, while no peaks appear in the current of defected SiNRs which increased almost linearly at the bias range of 0.4V-1V. To show the asymmetry property of the system, the rectification ratios were computed. The rectification ratios is defined as R(V)=|I(V)/I(-V)| which is shown in Fig.4. The R(V)=1 indicates no rectification. R(V)>1 indicates that the current in the positive voltage is higher than that in the negative voltage. From the rectification ratio of SW-defected SiNRs which illustrated in Fig.4, one can see that the rectification ratio remains larger than 1.0 after 0.65V and comes to the largest value of about 1.4 at 1.0V, which is larger than that of the perfect SiNRs. The results show that SW-defected SiNRs device possess better rectification effect at high bias than perfect SiNRs does.

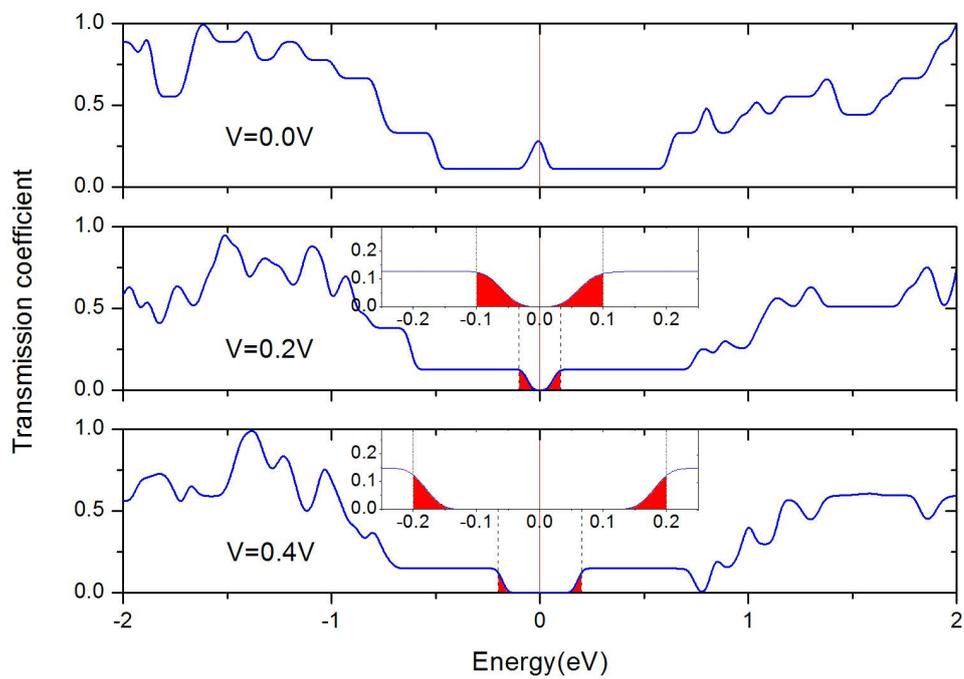

(a)

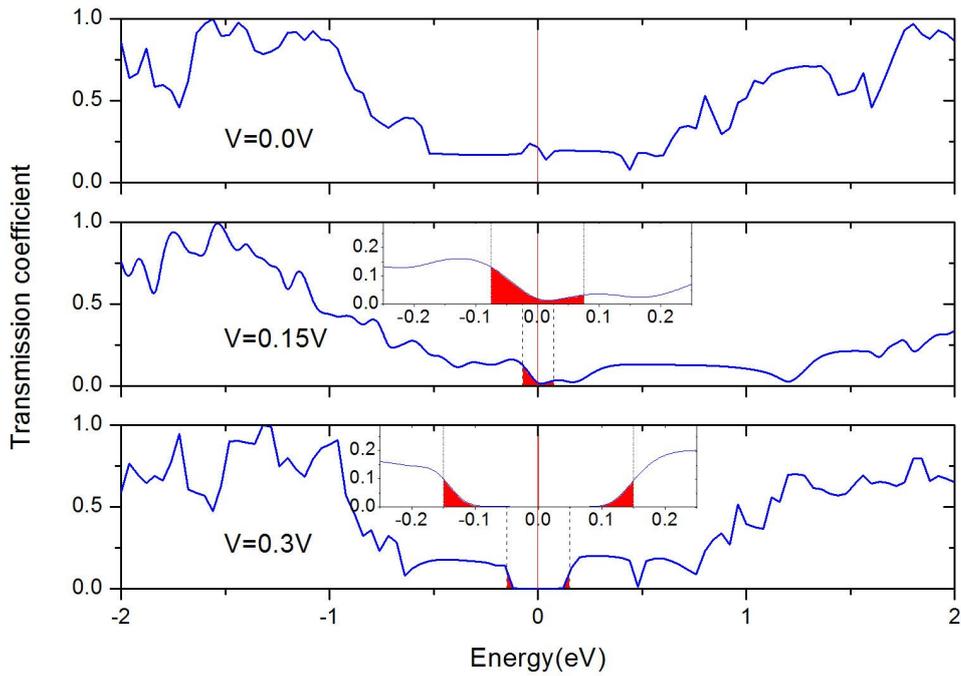

(b)

Fig. 4 Transmission spectra of perfect SiNRs (a) and SW-defected SiNRs (b) under different applied bias. The filled region represents the bias window. The inset shows the local expansion for lower bias range.

Fig. 4 summarizes the transmission spectra for the perfect SiNRs and SW-defected SiNRs under different bias. The average Fermi level is set as zero and the bias window which contributes to the current integral is indicated by the filled region. Since the current through the system is an integration of the transmission coefficient within the bias window around the Fermi level, only a finite part of the transmission spectrum should be analyzed. For perfect SiNRs, by increasing the bias voltage to 0.2 V, the current increases while new transmission enters the bias voltage window. When the bias voltage attains to 0.4 V, the transmission spectrum across the bias window decreases and cannot be compensated by the increase of bias window, which results in the negative differential resistance (NDR) behavior in I-V curves in Fig.3. For SW-defected SiNRs, the NDR behavior appears in the range from 0.15V to 0.3V, narrower than that of the perfect SiNRs.

In order to further understand the NDR behavior, Table 1 shows the MPSH of the highest occupied molecular orbital (HOMO), the lowest unoccupied molecular orbital (LUMO), and their nearby orbitals HOMO-1 and LUMO+1. Different bias values have been considered for SW-defected SiNRs (0.2v and 0.4v), perfect SiNRs (0.15v and 0.3v). We note that the MPSH of LUMO and LUMO+1 are delocalized and the corresponding transport channels are opened for perfect SiNRs at 0.2V. LUMO+1 is localized at 0.4V, makes little contribution to the charge transport. So we can find the NDR behavior in I-V curve. For SW-defected SiNRs, the MPSH of LUMO and LUMO+1 are delocalized at 0.15V, LUMO is localized at 0.3V. The similar phenomenon also appears in SW-defected SiNRs. Note that if the orbit is delocalized across the molecule, an electron entering the molecule at the energy of the orbital has a high probability of going through the molecule [25]. So we can know that the frontier molecular are opened for electronic transport at a lower bias in a voltage range, while they are suppressed at a higher bias, resulting in the observation of NDR behavior in I-V curves in Fig. 4.

| Model | Bias | HOMO-1 | HOMO | LUMO | LUMO+1 |
|---|---|---|---|---|---|
| perfect SiNRs | 0.2 | 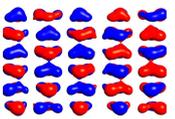 | 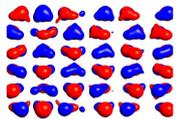 | 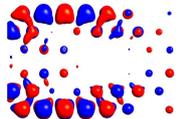 | 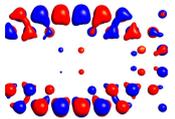 |
| | 0.4 | 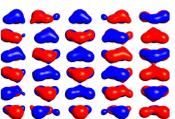 | 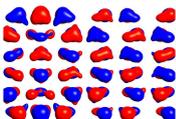 | 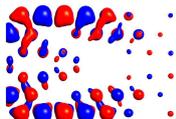 | 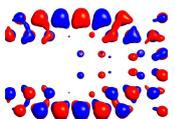 |
| SW-defected SiNRs | 0.15 | 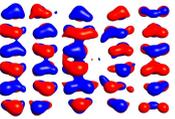 | 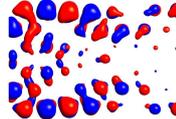 | 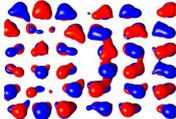 | 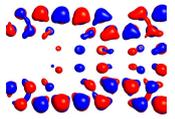 |
| | 0.3 | 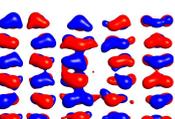 | 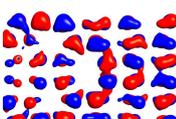 | 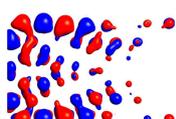 | 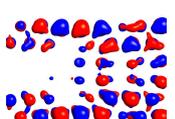 |

Table 1 Frontier molecular orbitals of MPSH for SW defected SiNRs at specific bias voltages.

## 5. Conclusion

In summary, we have investigated the transport properties of the molecular junctions based on the structural and transport properties of zigzag graphene nanoribbons with Stone-Wales defect using DFT combined with the NEGF formalism. The calculated formation energy is significantly lower than that of graphene and silicene, implies the high stability of such defect in SiNRs. The NDR behaviors are observed within a certain bias voltage range for both the perfect SiNRs and SW-defected

SiNRs, while SW-defected SiNRs device possess better rectification effect at high bias than perfect SiNRs does. Our calculation results suggested that the NDR behavior originates from the suppression of the frontier molecular orbitals with the bias increasing.

**Acknowledgements:**

The authors would like to acknowledge the support by the Fundamental Research Funds for the Central Universities of China, No: 2012-Ia-051.